\documentclass[12pt]{article}
\usepackage{fullpage}
\usepackage{amsmath}
\usepackage[margin=1.5cm]{geometry}
\usepackage{graphicx}

\title{\flushleft{\scshape{Review Article}}\\ \center\textbf{Quantum nanophotonics using hyperbolic metamaterials}}
\author{\textbf{C. L. Cortes, W. Newman, S. Molesky and Z. Jacob} \\ 
Department of Electrical and Computer Engineering, University of Alberta,\\
Edmonton, AB T6G 2V4, Canada \\
\\
E-mail: zjacob@ualberta.ca}
\date{}
\begin{document}

\maketitle

\begin{abstract}
Engineering the optical properties using artificial nanostructured media known as metamaterials has led to breakthrough devices with capabilities from super-resolution imaging to invisibility. In this article, we review metamaterials for quantum nanophotonic applications, a recent development in the field. This seeks to address many challenges in the field of quantum optics using recent advances in nanophotonics and nanofabrication. We focus on the class of nanostructured media with hyperbolic dispersion that have emerged as one of the most promising metamaterials with a multitude of practical applications from subwavelength imaging, nanoscale waveguiding, biosensing to nonlinear switching. We present the various design and characterization principles of hyperbolic metamaterials and explain the most important property of such media: a broadband enhancement in the electromagnetic density of states. We review several recent experiments that have explored this phenomenon using spontaneous emission from dye molecules and quantum dots.  We finally point to future applications of hyperbolic metamaterials of using the broadband enhancement in the spontaneous emission to construct single photon sources.
\end{abstract}

\section{Introduction}\label{introduction}  
The search for new materials naturally led to the concept of a metamaterial where the bulk electromagnetic properties arise not only due to the composition but also because of the underlying structural resonances and near field coupling between the designed subwavelength building blocks \cite{shalaev2007optical}. The simultaneous progress in the field of nanofabrication has been a major impetus for this unprecedented control over the microscopic structure. The availability of metamaterials led to exotic applications such as negative refraction \cite{foteinopoulou2003refraction}, subwavelength resolution imaging in the near and far field \cite{fang2005sub--diffraction-limited,jacob2006optical,liu2007far-field,liu2007far-fielda,smolyaninov2007magnifying}, invisibility devices \cite{pendry2006controlling} and perfect absorbers \cite{landy2008perfect}.

The class of artificial media which has emerged as one of the most important at optical frequencies are hyperbolic metamaterials (HMMs). These non-magnetic media have a dielectric tensor which is extremely anisotropic and cannot be found in nature at optical frequencies. The name is derived from the isofrequency curve of the medium which is hyperbolic as opposed to circular as in conventional media.  The work in optical hyperbolic metamaterials was initiated by their ability to exhibit negative refraction \cite{smith2004negative} and index \cite{podolskiy2005strongly} but since then, it has diversified into a multitude of applications from  subwavelength imaging \cite{jacob2006optical}, nanoscale waveguiding and light confinement \cite{govyadinov2006metamaterial, yao2011three}, biosensing \cite{kabashin2009plasmonic} to enhanced nonlinearities \cite{wurtz2011designed}. The primary reason for this is the  highest figure of merit amongst all other metamaterials, ease of nanofabrication, bulk three dimensional non-resonant response, deeply subwavelength unit cells and tunability across a broad range of wavelengths \cite{hoffman2007negative}.

Historically, hyperbolic media were first explored in the field of plasma physics \cite{felsen1964focusing}. By applying a strong magnetic field to a plasma, the dielectric tensor could be made extremely anisotropic leading to hyperbolic isofrequency curves. In fact,  negative refraction and planar focusing due to hyperbolic dispersion were predicted in anisotropic plasmas \cite{felsen1964focusing} before Veselago's seminal paper on the negative index lens \cite{veselago1968electrodynamics}. Other characteristic features such as the nature of dipole radiation and propagation in hyperbolic media were also addressed by the pioneering work of L. Felsen \cite{felsen1994radiation}. Inspired by the work in plasmas, low frequency metamaterials were designed to emulate many of the hyperbolic properties using microwave circuits that can be ascribed an effective index \cite{balmain2002resonance}.

This article will focus on hyperbolic metamaterials at mid-infrared (mid-IR) and optical wavelength ranges. The range of applications at high frequencies such as subwavelength imaging or nanowaveguiding are fundamentally different and so too are the challenges. The purpose of this article is two-fold. First, we provide a comprehensive design perspective on hyperbolic metamaterials from the visible to mid-infrared wavelengths. We compare and explain the choice of materials and design parameters which have been used for various experiments. This should help experimentalists design hyperbolic metamaterials for specific device applications. We focus extensively on the validity of the effective medium approximation to describe hyperbolic metamaterials.  The underlying theme of the review is the emerging area for plasmonics and metamaterials research: quantum applications \cite{jacob2011plasmonics}.

It was recently predicted by Z. Jacob \emph{et al} that hyperbolic metamaterials have a large photonic density of states leading to a broadband Purcell effect useful for applications such as efficient single photon sources \cite{jacob2009broadband,jacob2009single,jacob2010classical}. This was later experimentally \cite{jacob2010engineering,noginov2010controlling,krishnamoorthy2012topological} and computationally \cite{kidwai2011dipole,{poddubny2011spontaneous}} verified by multiple groups, proving that hyperbolic metamaterials are a viable route for radiative decay engineering. The metamaterial approach of modifying spontaneous emission is fundamentally different from methods that involve resonant cavities \cite{michler2000quantum,{pelton2002efficient}}, photonic crystal structures \cite{lodahl2004controlling,{hughes2004enhanced}}, slow light waveguides \cite{yao2009ultrahigh} or low mode volume plasmons \cite{chang2006quantum}. The broadband enhancement in the spontaneous emission occurs due to coupling with electromagnetic states unique to the hyperbolic metamaterial. We review in detail the various experiments that probed the predicted effect and compare it with the theoretical results \cite{hoffman2007negative,noginov2009bulk}.

The main application of metamaterial based quantum nanophotonics would involve the coupling to quantum emitters such as nitrogen-vacancy centers in diamond \cite{{jacob2011plasmonics},babinec2010diamond}. Their broadband emission are not ideally suited to the optical cavity based approach for enhancement. In spite of the absorption losses, an optimized design can lead to efficient single photon sources based on hyperbolic metamaterials \cite{jacob2009broadband,{lounis2005single-photon}}. The issue of spontaneous emission near a hyperbolic metamaterial is also pertinent to the problem of loss compensation. A large number of emitters would emit preferentially into metamaterial modes leading to active photonic devices. Finally, we expect this review to pave the way for future work on non-classical light interaction with metamaterials.

\begin{figure}[t]	
\centering
\includegraphics[width=160mm]{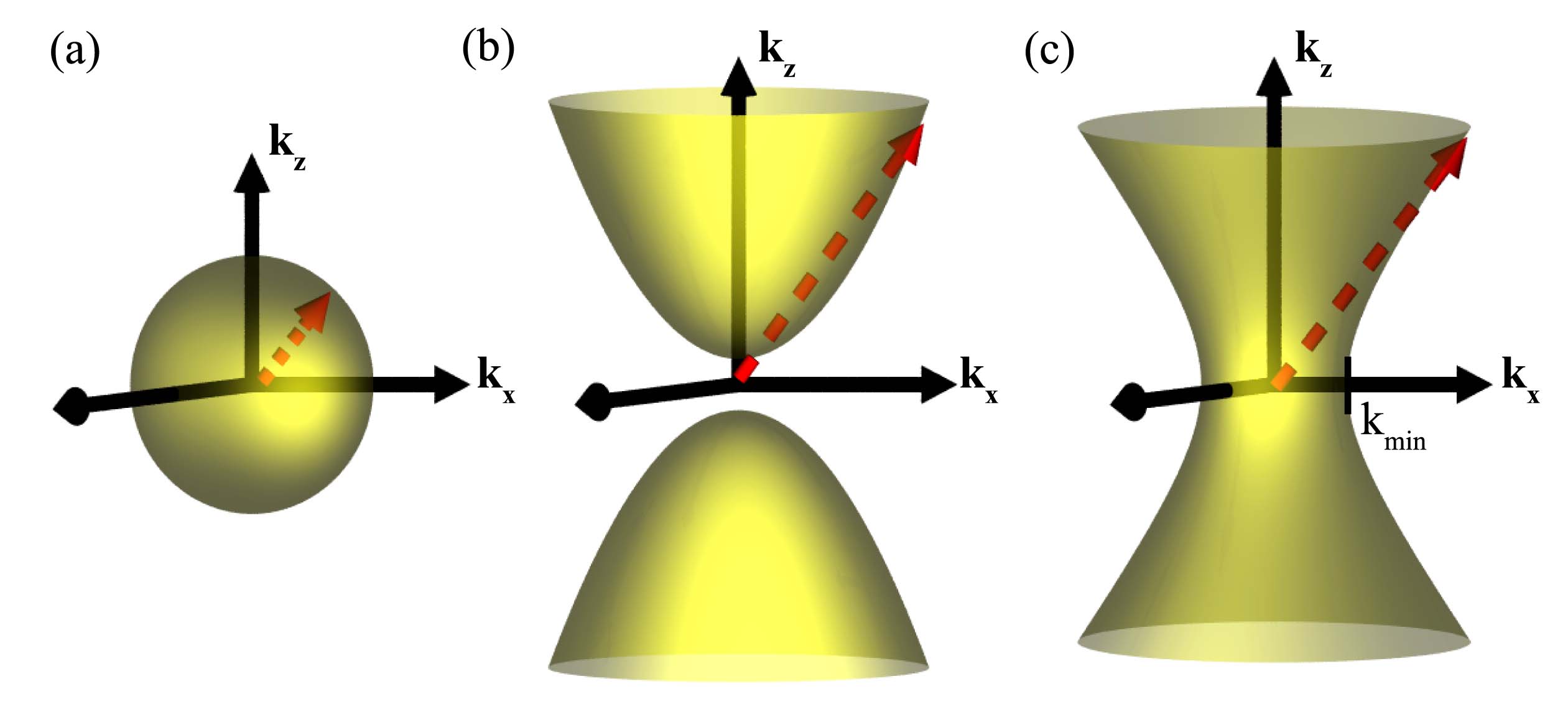}
\caption{\emph{k}-space topology. The isofrequency contour for  (a) an isotropic dielectric is a sphere, and for (b) extraordinary waves in a uniaxial medium with extreme anisotropy ($\epsilon_x=\epsilon_y > 0$ and $\epsilon_z < 0$) it is a hyperboloid (type I HMM). (c) Hyperboloid of a type II metamaterial when two 
components of the dielectric tensor are negative ($\epsilon_x=\epsilon_y < 0$ and $\epsilon_z > 0$). This medium behaves like a metal reflecting all waves below a minimum parallel wavevector $k_{min}$. The 
metamaterials in (b) and (c) can support waves with unbounded wavevectors (dashed red arrow) 
as opposed to an isotropic medium.  }
\label{topology}
\end{figure}

\section{Design and Characterization of the HMM}\label{design}
\subsection{\it{Isofrequency relation}}
The fundamental property of the HMM is that it has dielectric properties ($\epsilon>0$) in one direction but metallic properties ($\epsilon<0$) in an orthogonal direction. The dielectric response of the metamaterial is that of a uniaxial crystal ($\epsilon_{xx}=\epsilon_{yy}\neq\epsilon_{zz}$)
\begin{equation}
\epsilon=
\left[ \begin{array}{ccc}
  \epsilon_{xx} & 0 &  0 \\
  0 & \epsilon_{yy} &  0 \\
  0     & 0     & \epsilon_{zz}
 \end{array} \right]
\label{uniaxial_tensor}
\end{equation}
but with extreme anisotropy $\epsilon_{xx}\cdot \epsilon_{zz} < 0$. Such a property does not exist in nature at optical frequencies and has to be engineered using nanostructured metamaterials. The word hyperbolic metamaterial arises from the unique isofrequency surface of extraordinary waves in such a medium which is given by
\begin{equation}
\frac{k_x^2 + k_y^2}{\epsilon_{zz}} + \frac{k_z^2}{\epsilon_{xx}} = \frac{\omega^2}{c^2}
\label{dispersion_relation}
\end{equation}
Note that the isofrequency surface is simply a sphere, as seen in \ref{topology} (a), for an isotropic medium  ($\epsilon_{xx}=\epsilon_{yy}=\epsilon_{zz}$), but opens into a hyperboloid when the dielectric constants are of opposite signs along the principle axes. Two possibilities exist depending on which direction is metallic leading to either  a single-sheeted hyperboloid or a double-sheeted hyperboloid (see \ref{topology} (b) $\&$ (c)). We classify these two systems based on the number of negative components of the dielectric tensor as type I ($\epsilon_{xx}=\epsilon_{yy} > 0$ while $\epsilon_{zz}<0$) and type II  ($\epsilon_{xx}=\epsilon_{yy} < 0$ while $\epsilon_{zz}>0$) hyperbolic metamaterials.

The most striking feature of such metamaterials, evident from the isofrequency surface, is the existence of unique metamaterial states which can have wavevectors far exceeding the free space wavevector ($k_0=\omega/c$). They are referred to as the ``high-\emph{k}" states of a hyperbolic metamaterial  and lie at the heart of many device applications from imaging to quantum nanophotonics. These high-\emph{k} states are evanescent in vacuum and decay away exponentially. In stark contrast, these waves are allowed to propagate within an HMM. The properties and applications of these metamaterial states are discussed in Section 3 and 4 respectively and we now focus on practical designs which can achieve the desired electromagnetic response.
\subsection{\it{Practical designs and materials perspective}}
\begin{figure}[t]	
\centering
\includegraphics[width=160mm]{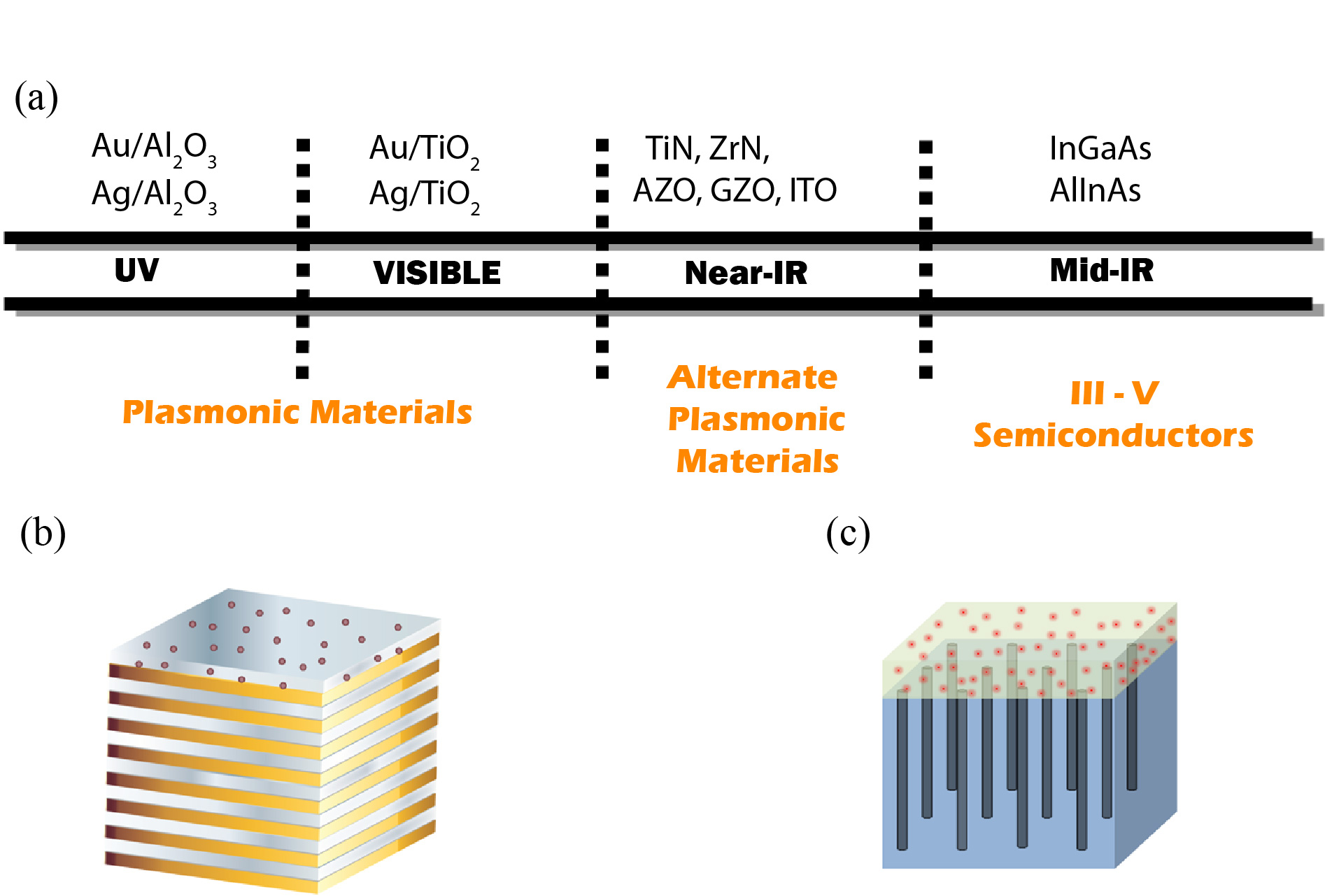}
\caption{
Hyperbolic media: A materials perspective (a) Hyperbolic metamaterials can be made by a 
variety of plasmonic materials tailored to different regions of the electromagnetic spectrum from 
the visible to mid-infrared frequencies (b) multilayer realization consists of alternating 
subwavelength layers of metal and dielectric (c) HMM based on metal nanowires in a dielectric host. The spontaneous emission from dye molecules near these metamaterials were studied in references \cite{krishnamoorthy2012topological,kidwai2011dipole}.}
\label{materials}
\end{figure}

There are two main approaches to achieving the desired hyperbolic isofrequency surface using metamaterials. The first one relies on a metal-dielectric superlattice with subwavelength layer thicknesses \cite{jacob2006optical,liu2007far-field,xiong2007two-dimensional,hoffman2007negative}; shown in \ref{materials} (b). The second approach utilizes metallic nanowires in a dielectric host \cite{noginov2009bulk,elser2006nanowire,yao2008optical,kanungo2010experimental,pollard2009optical,casse2010super-resolution} as seen in \ref{materials} (c). Effective medium theory (see appendix A and B) shows that these two structures can achieve the required extreme anisotropy. It is well known that the dielectric constants of nanostructured metal, inevitable in hyperbolic metamaterial designs, are different from the established values of their bulk counterparts \cite{Ni2008,johnson1972optical}. Higher absorption in the metallic thin films and nanowires poses a major challenge for hyperbolic media and metamaterials in general. Deposition of smooth thin films of silver (for the multilayer design) with low loss can be achieved using a germanium wetting layer \cite{chaturvedi2010smooth,chen2010ultra}; while the quality of thin metallic rods (for the nanowire design) has been shown to be significantly improved by annealing \cite{pollard2009optical}. In fact, hyperbolic metamaterials have the highest figure of merit among any demonstrated metamaterials. 


\begin{figure}[t]	 
\begin{minipage}[b]{0.5\linewidth}
\centering
\includegraphics[width=80mm]{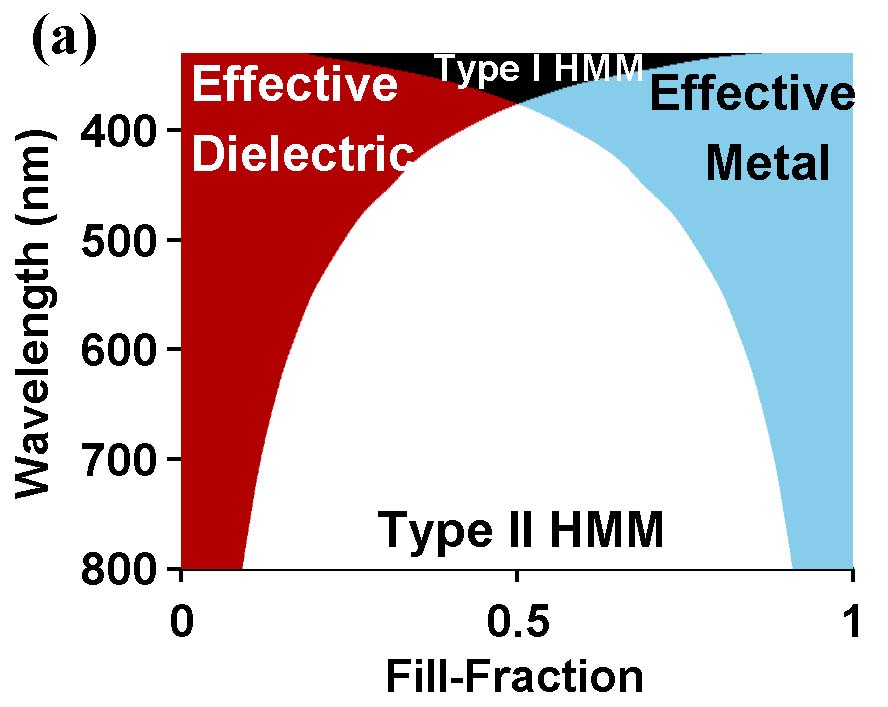}
\end{minipage}
\hspace{0.1cm}
\begin{minipage}[b]{0.5\linewidth}
\centering
\includegraphics[width=80mm]{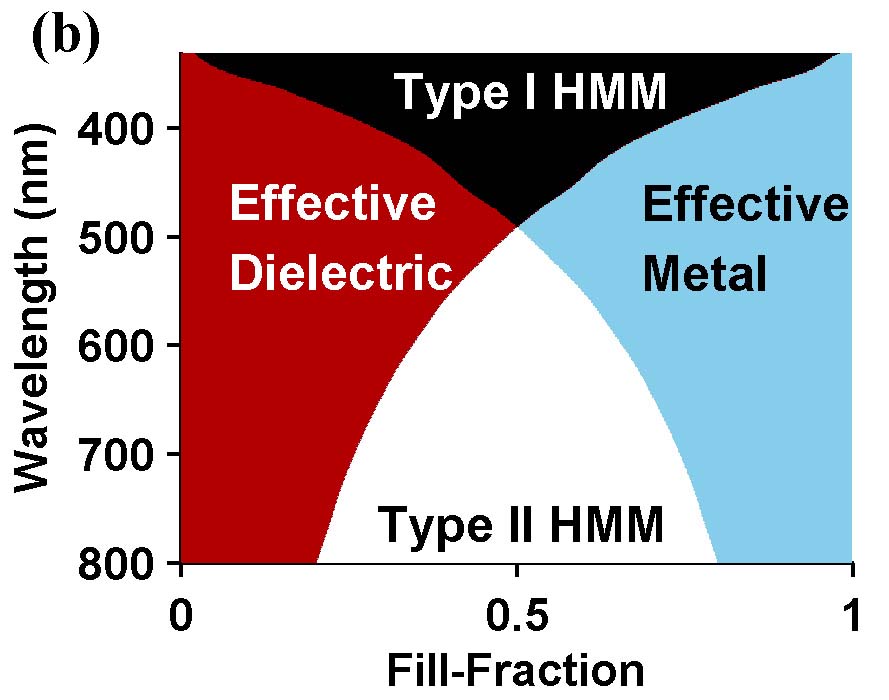}
\end{minipage}
\begin{minipage}[b]{0.5\linewidth}
\centering
\includegraphics[width=80mm]{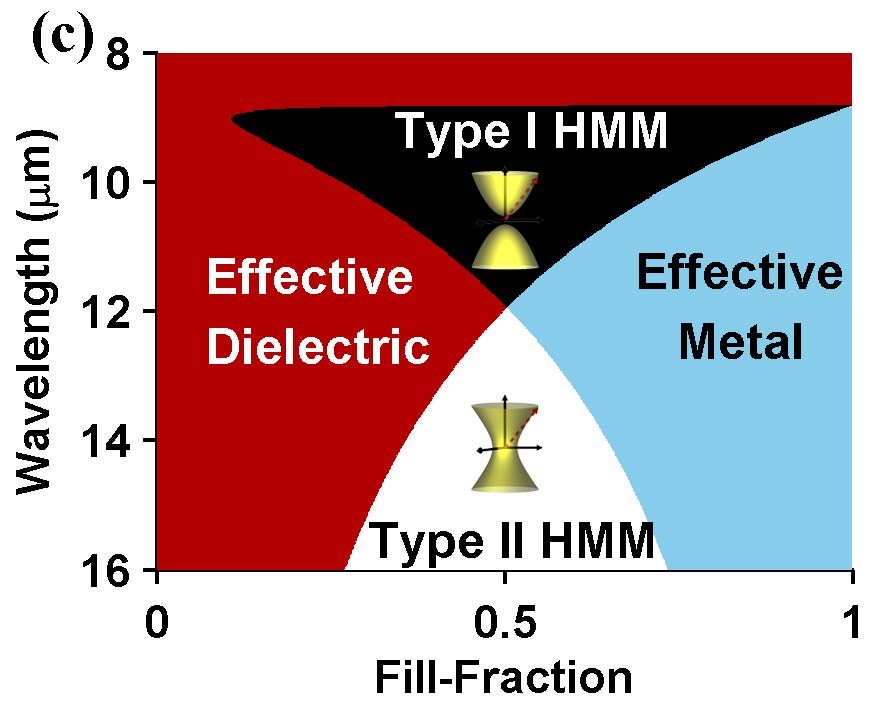}
\end{minipage}
\hspace{0.1cm}
\begin{minipage}[b]{0.5\linewidth}
\centering
\includegraphics[width=80mm]{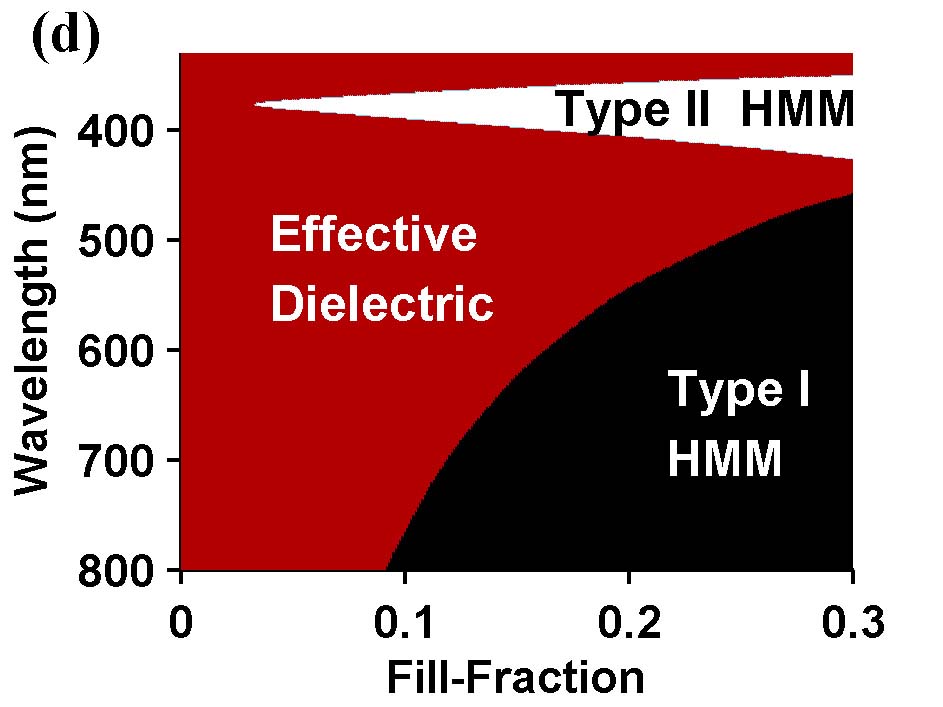}
\end{minipage}
\caption{
EMT predicts that a metal dielectric composite can behave as an effective dielectric, effective 
metal, type I HMM or type II HMM depending on the wavelength and fill fraction of the metal. 
Optical phase diagrams for (a) $Ag/Al_2O_3$ multilayer system (b) $Ag/TiO_2$ multilayer system (c) 
AlInAs/InGaAs multilayer system in the mid-IR region (d) Silver nanowires in an alumina 
matrix.}
\label{opd}
\end{figure}

\begin{figure}[t]	
  \centering
  \begin{minipage}{8cm}
    \includegraphics[width=80mm]{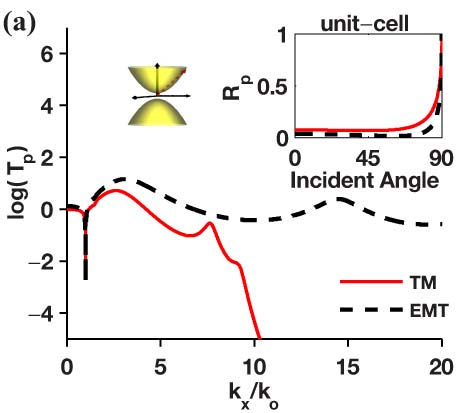}  
  \end{minipage}
  \begin{minipage}{7.5cm}
    \includegraphics[width=80mm]{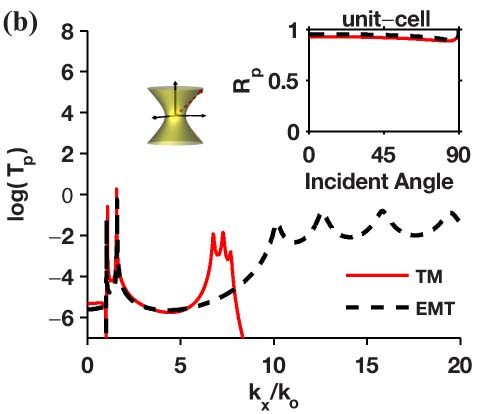}  
  \end{minipage}
\caption{Comparison of EMT and the transfer matrix (TM) method. The transmission is shown (a) at 440 nm, for a type I HMM consisting of a 6-layer $Ag/TiO_2$ system (15 nm each) (b) at 750 nm, for a type II HMM with 0.7 fill-fraction (35nm/15nm). The insets show the reflectivity for the corresponding unit cells.}
\label{r_and_t}
\end{figure}

The above mentioned designs can be tuned to function in all wavelength ranges of interest from the UV, visible, near-IR to mid-IR with the appropriate choice of metal and metallic filling fraction. We note that the key design aspects are the choice of the metal and its plasmonic response. In the visible range, silver forms the best choice because of its low optical losses, however as the wavelength increases we need to resort to other metals in the near-infrared (telecommunication wavelength) and mid-infrared regions. This is primarily because of the large reflectivity of the metal at higher wavelengths (lower frequencies) which leads to a large impedance mismatch with other media and thus reduces transmission. The best alternative in the long-wavelength region is a highly doped semiconductor which can behave as the plasmonic building block for hyperbolic metamaterials \cite{hoffman2007negative}. The doping controls the plasma frequency of the metal and for III-V semiconductors can be tuned up to the mid-IR. At the near-IR wavelength ranges, transparent conducting oxides and transition-metal nitrides have emerged as an alternative plasmonic material to silver \cite{boltasseva2011low-loss,{west2010searching}}. In figure \ref{materials} (a), we outline different materials that can be used as building blocks for hyperbolic metamaterials.

For the choice of dielectric, a high refractive index is required to achieve a type I response as well as to get a high throughput. One of the options is titanium dioxide ($TiO_2$) in the visible spectrum which has a high dielectric constant. Close to UV wavelengths, however, $TiO_2$ has high absorption and so alumina ($Al_2O_3$) becomes a better alternative.  In figure \ref{opd}, we show the optical phase diagram for the hyperbolic metamaterial where the effective medium response is plotted with the wavelength and the fill-fraction of metal. We consider both the multilayer and nanowire designs as well as visible and mid-IR wavelength ranges. The design and operating point for specific applications can be chosen from this graph. We note that all structures show both type I and type II hyperbolic metamaterial behavior in a broadband wavelength region. This is primarily because of the non-resonant nature of the desired electromagnetic response. The homogenization also predicts this metal-dielectric mixture to behave like an effective metal or effective dielectric  where interesting properties are not expected. The transitions between the various surfaces in \emph{k}-space occur within the borders of the different coloured regions in the graphs \cite{krishnamoorthy2012topological}. The transition point where the four effective materials meet represents the plasmon resonance of the metal-dielectric metamaterial composite. 

\subsection{\it{How effective is effective medium theory (EMT)?}}
A natural question arises whether the optical phases and hyperbolic isofrequency surfaces predicted by effective medium theory are achievable in the practical structure. Homogenization is only valid when the wavelength of light is much larger than the unit cell of the metamaterial. In hyperbolic metamaterials, the interesting properties arise due to propagating waves that have an effective wavelength inside the medium which is much smaller than the free space wavelength (high-\emph{k} states). We therefore compare the results of effective medium theory and a practical multilayer realization of the hyperbolic metamaterial to validate the predictions. Throughout the review, we consider the effect of absorption, dispersion, unit cell size, total sample size and also the role of the substrate. Our conclusions apply for applications in imaging as well as quantum optics and, therefore, should guide experimentalists to choose the optimum design for an experiment.


\begin{figure}[t]	
  \centering
  \begin{minipage}{8cm}
    \includegraphics[width=80mm]{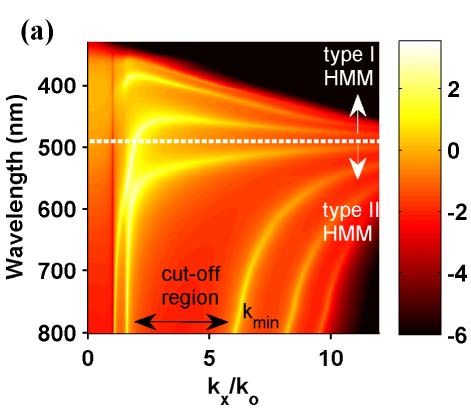}  
  \end{minipage}
  \begin{minipage}{7.5cm}
    \includegraphics[width=80mm]{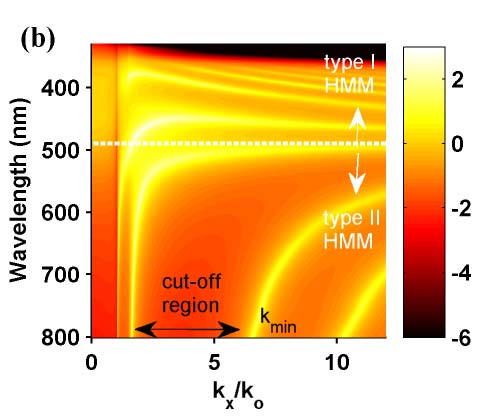}  
  \end{minipage}
\caption{
Comparison of the transmission of (p)-polarized propagating and evanescent waves through the 
metamaterial calculated using (a) transfer matrix for an 8-layer system consisting of $Ag$ and $TiO_2$ (15 nm each) and (b) effective medium theory. The colour bar corresponds to the transmittance in log scale. The plasmon resonance occurs at a wavelength of $\lambda_p$ = 490 nm. The multilayer structure achieves a type I response for $\lambda<\lambda_p$ and a type II response for $\lambda>\lambda_p$. The existence of multiple high-\emph{k} metamaterial states away from the plasmon resonance is seen in the multilayer realization, in agreement with effective medium theory.}
\label{2Dtransmission}
\end{figure}

\subsubsection{\it{Propagating and evanescent wave spectrum:}}
\begin{figure}[t]	
\begin{minipage}[b]{0.5\linewidth}
\centering
\includegraphics[width=80mm]{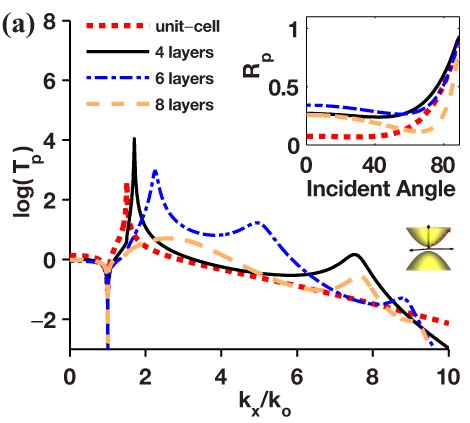}
\end{minipage}
\begin{minipage}[b]{0.5\linewidth}
\centering
\includegraphics[width=80mm]{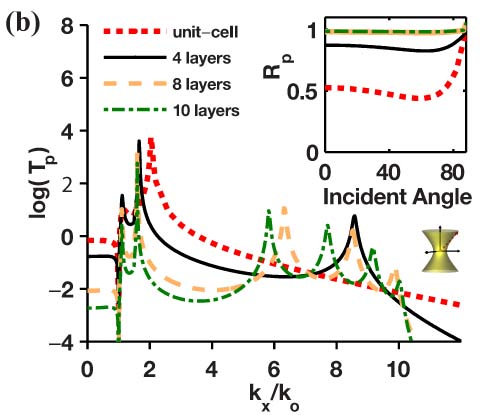}
\end{minipage}
\caption{
Transmission dependence on the number of layers of a multilayer HMM (Ag/$TiO_2$ 15 nm each).  (a) Type I metamaterial ($\lambda$ = 440 nm) for varying number of layers. Inset: Low reflectivity 
characteristic of type I HMM which has ($\epsilon_{||} = \epsilon_x = \epsilon_y > 0$). (b) Type II HMM 
($\lambda$ = 750 nm) which shows the presence of surface plasmon polaritons as well as high-\emph{k} 
modes. Inset: Reflectivity closely resembles that of a metal. The $k_{min}$ value of 4.2$k_0$ agrees 
with EMT.  Notice that the unit cell in both cases shows no 
high-\emph{k} states unlike the multilayer structures.}
\label{nolayers}
\end{figure}

Routine characterization of metamaterials proceeds by measurement of both the phase and amplitude of the reflected light. Ellipsometric characterization along with the knowledge of the uniaxial nature of the multilayer or nanowire structure can uniquely identify the components of the dielectric tensor. We note that type I HMMs always have lower reflections than type II because of the positive tangential component of the dielectric constant. Comparing the reflected spectrum of propagating waves in the insets of figure \ref{nolayers}, we see a good match between the effective medium predictions and the practical multilayer structure for both type I and type II metamaterials. Irrespective of the fill-fraction, excellent agreement is obtained as long as the unit cell is significantly subwavelength. Surprisingly, even the unit cell reflectivity starts agreeing with the EMT prediction (figure 4 inset), but this is not always the case (figure 6 inset).

Propagating waves can only probe the ``low-\emph{k}'' spectrum of waves allowed within the HMM. Since the most interesting properties stem from high-\emph{k} propagating waves within the metamaterial, it becomes necessary to experimentally and theoretically understand the tunneling of evanescent waves through the multilayer structure. In figure \ref{r_and_t}, we note the presence of multiple peaks in the transmission of large wavevector waves which would normally be evanescent and simply decay away in any dielectric medium. These correspond to coupled surface plasmon polariton waves propagating on the thin metallic surfaces described in the effective medium theory as high-\emph{k} states. Deviations from effective medium theory are natural and expected for incident evanescent waves because the effective wavelength of such waves starts approaching the unit cell size. Yet the key feature of high-\emph{k} modes predicted by EMT is clear in the practical multilayer structure. figure \ref{2Dtransmission} demonstrates a similar comparison between the multilayer metamaterial (a) and the effective medium prediction (b) across the optical wavelength range. We note that the high-\emph{k} modes in both the type I and type II metamaterials converge at the plasmon resonance wavelength. In addition, we see a noticeable cut-off region for the type II HMM in which no light is transmitted through the metamaterial. Transmission is allowed between the $k_{min}$ point and the unit cell cut-off point.

\subsubsection{\it{Number of layers:}}
The effective medium limit is normally attained only in the limit of a large collection of unit cells, ($N \to \infty$). We show that both the reflected and transmitted waves in the multilayer realization start matching the effective medium predictions with as little as a total of 8 layers. Starting with the unit cell, the addition of metal layers is seen to add more high-\emph{k} modes as predicted by EMT (see figure \ref{nolayers}). This is due to the coupled plasmonic states which increase in number on adding more metal layers. This emphasizes the fact that a small number of unit cells deviates significantly from the EMT prediction of the existence of high-\emph{k} waves. In the ideal limit, EMT predicts an infinite number of high-\emph{k} modes, because waves with arbitrarily large wavevectors can propagate in the hyperbolic metamaterial. However, in reality the finite size of the unit cell as well as the losses curtail the effect. Specifically, we see in figure \ref{nolayers} that the optimal transmission is obtained for the 6 layer system in the type I HMM, while the type II HMM shows optimal performance with a minimum of 8 layers. Type II HMM has pronounced high-\emph{k} peaks whereas the type I HMM has a flat transmission of high-\emph{k} states which makes it suitable for imaging applications.

\subsubsection{\it{Unit cell size:}}
\begin{figure}[t]	 
\begin{minipage}[b]{0.5\linewidth}
\centering
\includegraphics[width=80mm]{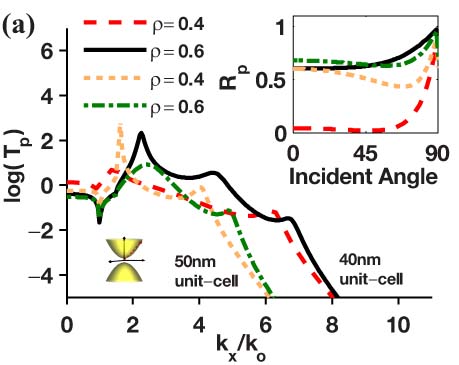}
\end{minipage}
\begin{minipage}[b]{0.5\linewidth}
\centering
\includegraphics[width=80mm]{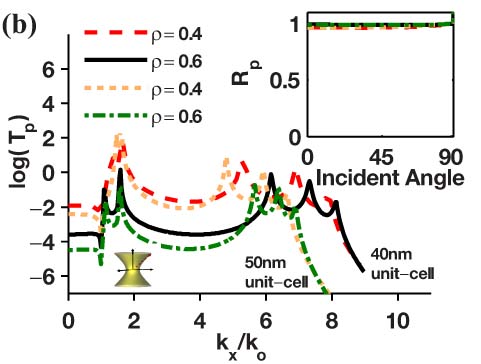}
\end{minipage}
\caption{A comparison of the p-polarized transmission for different fill-fractions. Using the transfer matrix method, the transmittance is calculated for a $Ag/TiO_2$ (a) 6-layer, type I HMM operating at 440 nm, and (b) an 8-layer type II HMM with an operating wavelength of 750 nm. The fill-fractions seen in the legend correspond to metal/dielectric thicknesses of 16nm/24nm, 24nm/16nm, 20nm/30nm, 30nm/20nm. The insets show the reflectance for p-polarized light.}
\label{fillfraction}
\end{figure}

A major limitation on the highest possible wavevector which propagates in the practical realization is the size, $a$, of the unit cell. We note that for waves inside the medium which have an effective wavevector $k_{eff} \sim 1/a$, the metamaterial limit is no longer valid. This is in agreement with the multilayer simulations which show that high wavevector transmission drops off considerably for $k_x \sim 1/a$. For such high-k waves, we enter the photonic crystal limit and the waves lie at the edge of the Brillouin zone of the 1D periodic lattice. In figure \ref{fillfraction}, we see excellent agreement between two different fill-fractions that have the same unit cell size and therefore have approximately the same cut-off point. Also note that as the unit cell size increases the cut-off point decreases as well.

\subsubsection{\it{Effect of loss:}}
Even with 20 to 30 unit cells consisting of silver and titanium dioxide, the large wavevector states are transmitted through the metamaterial \cite{hoffman2007negative}. Thus the biggest limitation to device performance arises from the unit cell size and the losses mainly effect the propagation length of the metamaterial states. This is the main reason hyperbolic metamaterials have a very high figure of merit and optimum device performance occurs away from a resonance where the bandwidth can be large and sensitivity to losses is lower.

\subsection{\it{Regime of operation}}

The wavelength and design for operation relies heavily on the application of choice. In imaging, where a throughput of both propagating and evanescent waves are required for high resolution, the type I metamaterial which has lower reflection performs best. In fact, impedance matching becomes of critical concern and often the regime of operation can be chosen by using $\epsilon_{xx}(\lambda_{op})=1$ in the type I region. The property of type II hyperbolic metamaterials to reflect propagating waves and transmit only evanescent waves has also been used in imaging to reduce the image distortion of subwavelength features that can be caused by propagating waves \cite{xiong2007two-dimensional}. Near the plasmon resonance of the multilayer structure $\epsilon_{xx} \to 0$ and $\epsilon_{zz} \to -\infty$, however, this is not the optimum functional wavelength due to losses and the large impedance mismatch with surrounding media \cite{belov2005canalization,belov2006subwavelength}. For the nanowire metamaterial, hyperbolic isofrequency surfaces can be achieved with very low metal fill fraction making it exceptionally useful for practical applications. A considerable body of work has focused on the $\epsilon_{zz} \approx 0$ region of the nanowire array. This is also known as the \emph{L} resonance since it occurs due to the coupling of the individual longitudinal plasmons of the metal nanowire. Applications such as ultrafast nonlinear switching as well as effects like additional waves all occur near this \emph{L} resonance \cite{kabashin2009plasmonic,pollard2009optical}. For imaging, the type I hyperbolic behavior in the nanowire array has been demonstrated to yield excellent results both in the visible and near-IR (telecommunication wavelength) \cite{menon2008negative}. In section 4, we will elucidate the optimum regime of operation for quantum optical applications where a decreased lifetime of emitters and a high throughput are the desired figures of merit.

\subsection{\it{Extraction of effective medium parameters}}
 It is necessary to experimentally extract the dielectric tensor of a fabricated metamaterial to compare with the predictions of effective medium theory. This can be performed using just the amplitude of the (s) and (p) polarized light and a uniaxial model of the metamaterial. The s-polarized light has a circular isofrequency curve and the comparison of the measured reflectivity with the predicted value over a range of angles can lead to determination of the tangential component ($\epsilon_{xx}=\epsilon_{yy}$) uniquely at the specific wavelength of interest. The next step is to use this value and the reflected amplitude of p-polarized light for a range of angles to uniquely determine the third component of the dielectric tensor ($\epsilon_{zz}$).  A least-squares fitting procedure was used in \cite{jacob2010engineering,noginov2009bulk,tumkur2011control}, which gave good results. We emphasize that it is necessary in experiment to characterize the individual thin metallic layer and dielectric layer as the dielectric constants can be different than their bulk counterparts \cite{Ni2008,johnson1972optical}. A different approach is to study the extinction spectrum at different angles to characterize the anisotropy \cite{wurtz2011designed}. For nanowire samples, this helps to uniquely identify the $\epsilon_{zz}\approx 0$ wavelength which occurs at the \emph{L} resonance.  

To characterize the tunneling of evanescent waves and understand the deviations from effective medium theory, conventional spectroscopic ellipsometry is not sufficient. For hyperbolic metamaterials in particular, spatial dispersion place a critical role  since the characteristic high-\emph{k} metamaterial states have an effective wavelength in the medium comparable to the size of the unit cell. A total internal reflection set up can be used to measure the tunneling of evanescent waves and to characterize the dispersion of these unique states \cite{avrutsky2007highly}. The deviations from EMT are significant even for propagating waves at wavelengths for which $\epsilon_{zz}\approx 0$ \cite{pollard2009optical}. 

\section{Photonic Density of States}\label{pdos}
The striking property of the hyperbolic metamaterial which is key to understanding its quantum applications is the large number of electromagnetic states supported by it. The photonic density of states (PDOS) similar to its electronic counterpart is the means of quantifying this. An intuitive counting procedure in \emph{k}-space consists of calculating the volume between the isofrequency contours at $\omega(k)$ and $\omega(k) + \Delta \omega$. For vacuum, it is equivalent to the volume of an infinitesimally thin spherical shell in \emph{k}-space. As opposed to this, for the HMM we get a hyperboloidal shell which has infinite volume and in the effective medium limit leads to a broadband singularity in the density of states \cite{jacob2009broadband}. The physical reason behind this singularity is that there is no upper cut-off to the wavenumber in an ideal hyperbolic medium and infinitely large number of high-\emph{k} modes are supported by it.

\begin{figure}[t]	 
\begin{minipage}[b]{0.5\linewidth}
\centering
\includegraphics[width=80mm]{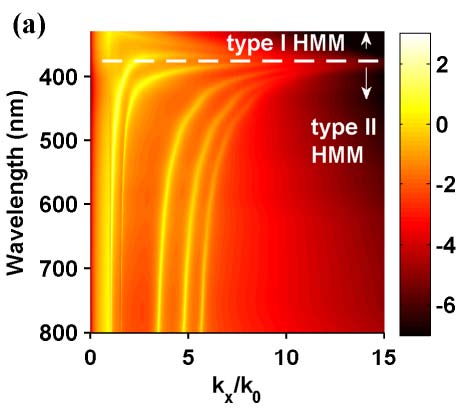}
\end{minipage}
\hspace{0.1cm}
\begin{minipage}[b]{0.5\linewidth}
\centering
\includegraphics[width=80mm]{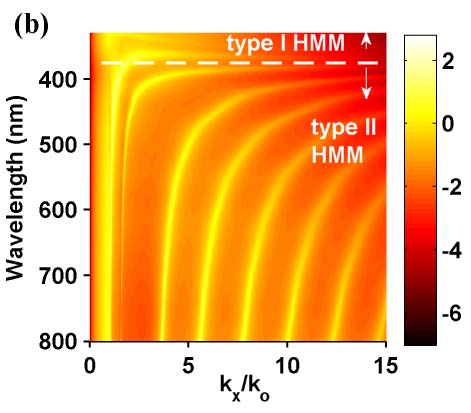}
\end{minipage}
\caption{W-LDOS in the near field of an HMM calculated by EMT and transfer matrix methods. The plots above correspond to (a) an 8-layer $Ag/Al_2O_3$, 15nm/15nm system calculated via the transfer matrix method and (b) the equivalent result predicted by EMT. The enhancement in the density of states in the multilayer system is curtailed by the finite unit cell size. }
\label{wldos}
\end{figure}

\subsection{\it{Wavevector-resolved Local Density of States (W-LDOS)}}
To quantify the above argument we need to define the density of electromagnetic states accessible in the near field of the metamaterial and also separate the contribution of the metamaterial modes which lead to the enhanced DOS. We therefore define the wavevector-resolved local density of states (W-LDOS) $\rho(\omega,d,\vec{k})$ which isolates the contribution to the LDOS of various modes (specified by the corresponding wavevector) at a distance \emph{d} in the near field of the metamaterial. The W-LDOS can be calculated in the near field of an effective medium slab with hyperbolic dispersion and compared with the corresponding multilayer realization. The W-LDOS is defined in terms of the Green's tensor for the multilayer or effective medium planar slab \cite{novotny2006principles,{ford1984electromagnetic}}. The explicit expression of the dyadic Green's function is given in appendix C. The total local density of states $\rho(r_o,\omega)$, at a frequency $\omega_o$, is
\begin{equation}
\rho(\mathbf{r_o}, \omega_o) = \frac{2\omega_o}{\pi c^2}\mathrm{Im}\{\mathrm{Tr}[\overleftrightarrow{G} (\mathbf{r_o},\mathbf{r_o}; \omega_o)]  \}.
\label{ldos}
\end{equation}
where $c$ is the speed of light.

We first consider a multilayer system consisting of $Ag/Al_2O_3$ and compare it with the predictions given by effective medium theory. In figure \ref{wldos} (a) and (b), the W-LDOS has been calculated for these two systems in the near field at a distance of 30 nm. The most interesting feature evident in the W-LDOS is the presence of peaks corresponding to the metamaterial modes with large wavevectors. Furthermore they are non-resonant in nature and are present in a broad bandwidth for both type I and type II metamaterials. As expected, in any practical realization, the enhancement in the LDOS is curtailed by the finite unit cell size. In spite of this, for deep subwavelength layers which are easily achievable by current nanofabrication techniques, an LDOS enhancement of over 20 is easily realizable. There is no contribution from propagating waves to the W-LDOS of the type II metamaterial (longer wavelengths) as opposed to the type I metamaterial (closer to the UV). A large increase in the LDOS is also seen closer to the surface plasmon resonance of the individual layers. In EMT, the parallel component of the dielectric tensor ($\epsilon_{x}$) passes through 0 at this wavelength. The losses are high at this resonance and it is better to function at a wavelength away from it.\\

\begin{figure}[t]	
\begin{minipage}[b]{0.5\linewidth}
\centering
\includegraphics[width=80mm]{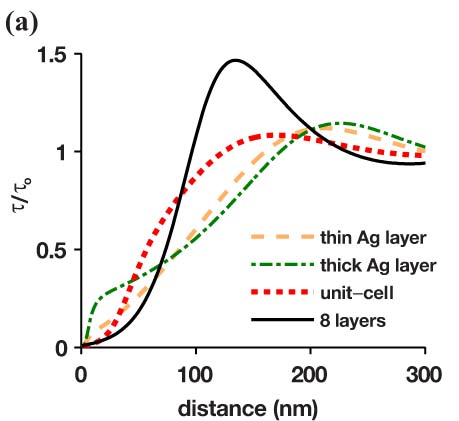}
\end{minipage}
\begin{minipage}[b]{0.5\linewidth}
\centering
\includegraphics[width=80mm]{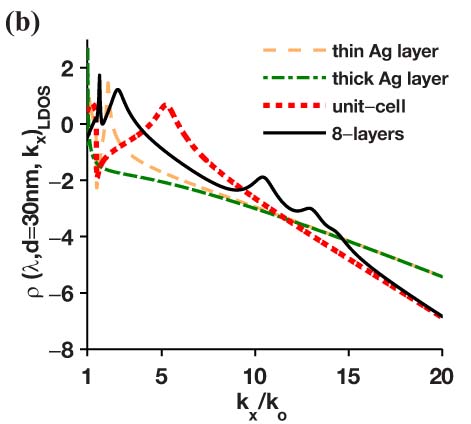}
\end{minipage}
\caption{(a) The lifetime in the near-field for a type II metamaterial with $\epsilon_{||}= -1.91+0.21i$ and $\epsilon_{\perp}=57.5 + 4.4i$ is shown to be lower than a thin or thick layer of silver. (b) The role of the high-\emph{k} states in the lifetime decrease is clear from the wavevector-resolved local density of states.}
\label{optimization}
\end{figure}
\section{Modification of the spontaneous emission lifetime}\label{lifetime}
Fermi's golden rule immediately shows that the enhanced local density of states has a consequence on the spontaneous emission lifetime of quantum emitters in the near field of the HMM. These emitters are best represented by point dipoles that emit both propagating and evanescent waves. In free space, excited fluorophores relax emitting energy only into propagating waves. But near the HMM, a new decay route arises due to coupling of the emitted evanescent waves to the high-\emph{k} metamaterial states. This significantly lowers the emitter lifetime. Furthermore, the dominant mechanism of relaxation in the near field is the emission into the multitude of metamaterial modes and not the propagating waves in vacuum.

The very presence of these large number of modes increases the decay rate causing an enhancement in the spontaneous emission which is known as the Purcell effect. Note that conventionally such enhancement has been studied near cavity resonances but these are fundamentally limited in bandwidth. In hyperbolic metamaterials, the enhancement is broadband and thus they can couple to emitters which have a broad spectral width.  Spontaneous emission lies at the heart of all light emitters from light emitting diodes, lasers to single photon sources. This is the main motivation behind many of the recent developments in engineering the spontaneous emission using hyperbolic metamaterials.

\subsection{\it{Coupling to metamaterial modes}}

In figure \ref{optimization}, we show the distance dependence of lifetime for a thin metal layer, the unit cell of the metamaterial and the multilayer hyperbolic metamaterial. The wavelength of operation is 545 nm, the thicknesses are 15 nm for both the metal and dielectric layers. We achieve a type II metamaterial and function near the transition region. In the near field, the lifetime on the multilayer structure is lowest in comparison to the others in agreement with effective medium theory. It should be noted that the lifetime decrease at close distances for the thin metal layer is due to the well known plasmonic route.  The local density of states for the three cases shows the presence of unique metamaterial states which is the primary reason for the lowest lifetime.
\begin{figure}[t]	
\begin{minipage}[b]{0.5\linewidth}
\centering
\includegraphics[width=80mm]{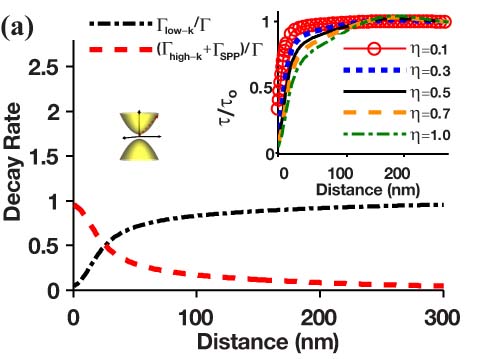}
\end{minipage}
\begin{minipage}[b]{0.5\linewidth}
\centering
\includegraphics[width=80mm]{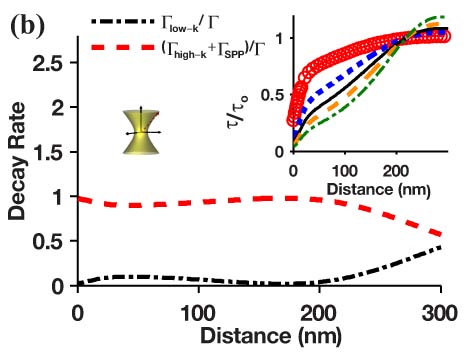}
\end{minipage}
\caption{
The coupling to the high-\emph{k} modes of a $Ag/TiO_2$ 8-layer metamaterial for (a) type I  ($\lambda$=440 nm) and (b) type II ($\lambda$=750 nm) metamaterial. Insets show the effect of varying intrinsic quantum yield of the emitter. Note that the near field decay rate is dominated by the metamaterial high-\emph{k} states. }
\label{highk_vs_lowk}
\end{figure}

The coupling to the metamaterial modes is due to the evanescent waves emitted by the point emitter and occurs only in the near field. We plot the decay rate contributions due to various routes of relaxation for the excited emitter in figure \ref{highk_vs_lowk}. The emission into propagating waves is reduced as the distance decreases while the opposite trend is noticed for the high-\emph{k} metamaterial states. The type II metamaterial, at longer visible wavelengths, also supports metamaterial surface plasmon polaritons which causes a  further reduction in lifetime. The insets shown in figure \ref{highk_vs_lowk} demonstrate the lifetime dependence on the intrinsic quantum yield $\eta$ of the emitter. As expected, small intrinsic quantum yield results in reduced coupling and a higher lifetime for any given distance of the dipole emitter.

\subsection{\it{Non-radiative decay rates}}
\begin{figure}[t]    
\begin{minipage}[b]{0.5\linewidth}
\centering
\includegraphics[width=80mm]{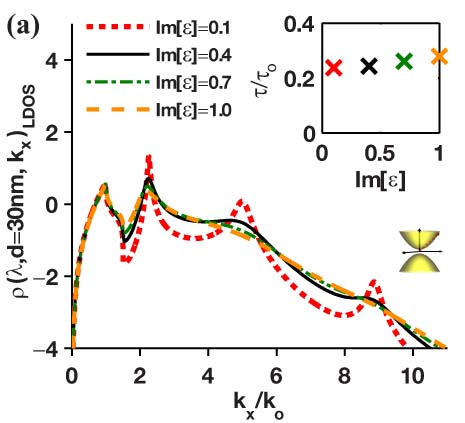}
\end{minipage}
\begin{minipage}[b]{0.5\linewidth}
\centering
\includegraphics[width=80mm]{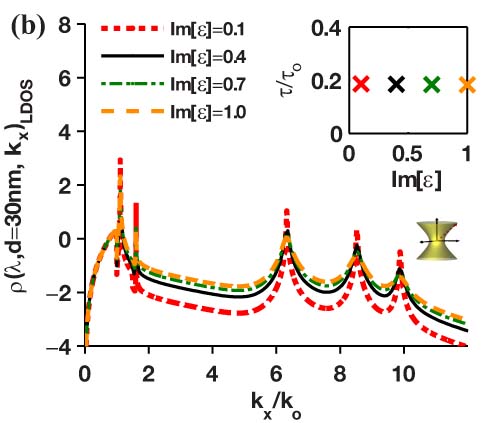}
\end{minipage}
\caption{The effect of losses on the wavevector-resolved local density of states, $\rho(\lambda,d,k_x)$ where d=30 nm. The area under the curve given $\rho(\lambda,d,k_x)$ determines the overall lifetime of the dipole emitter. The curves correspond to W-LDOS with different optical losses for (a) a 6-layer $Ag/TiO_2$ system at 440 nm (Type I HMM) and (b) an 8-layer $Ag/TiO_2$ at 750 nm (Type II HMM). Higher losses broadens the high-\emph{k} modes shown in the two figures; however, as shown in the insets, the lifetime $\tau/\tau_o$(normalized to vacuum) remains approximately the same.}
\label{losses}
\end{figure}

In figure \ref{losses}, we study the effect of losses on the local density of states. It is seen that the metamaterial modes which contribute to the LDOS broaden as the losses increase, corresponding to the decreased absorption length in the metamaterial. Yet the total LDOS (integrated over all wavevectors) is unchanged indicating that the role of losses is detrimental mainly to the length of propagation. This also shows that the enhancement in the LDOS is radiative in nature since even in the zero loss limit the metamaterial modes lead to the same LDOS enhancement.

\begin{figure}[t]	 
\begin{minipage}[b]{0.5\linewidth}
\centering
\includegraphics[width=80mm]{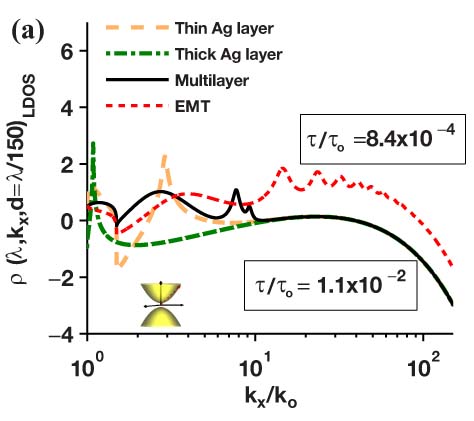}
\end{minipage}
\hspace{0.1cm}
\begin{minipage}[b]{0.5\linewidth}
\centering
\includegraphics[width=80mm]{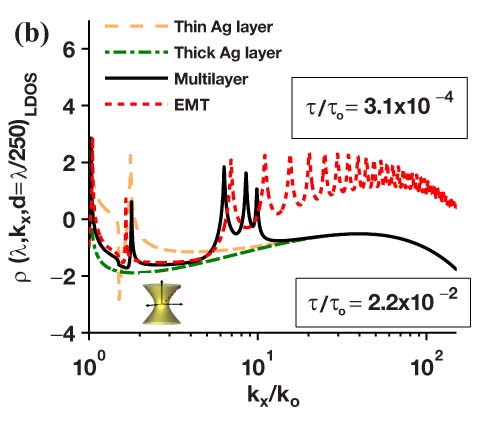}
\end{minipage}
\begin{minipage}[b]{0.5\linewidth}
\centering
\includegraphics[width=80mm]{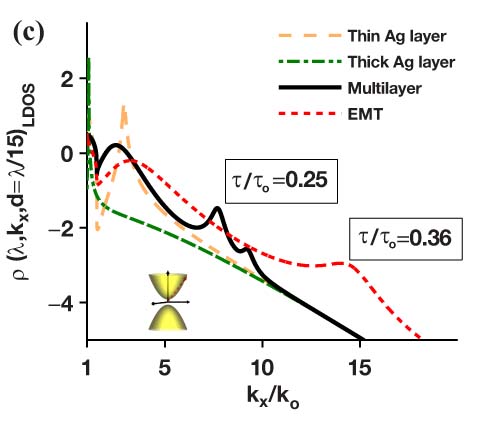}
\end{minipage}
\hspace{0.1cm}
\begin{minipage}[b]{0.5\linewidth}
\centering
\includegraphics[width=80mm]{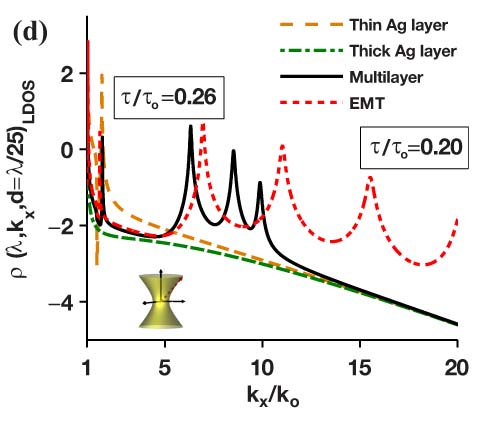}
\end{minipage}
\caption{
The effect of quenching for thick and thin silver layers, a multilayer system and an effective 
medium slab. The plots show the WLDOS for an emitter in the near-field. (a) $\lambda$=440 nm, $d=3$ nm (b) $\lambda$=750 nm and $d=3$ nm. The smooth peak in the WLDOS at very large wavevectors is indicative of quenching (lossy surface waves). The quenching is reduced at a distance of $d=30$ nm (c) $\lambda$=440 nm and (d) $\lambda$= 750 nm.}
\label{quenching}
\end{figure}

We note that a closer look at the quenching of spontaneous emission is necessary to decouple the radiative and non-radiative decay rates. The evanescent waves emanating from a point dipole in the near field of a lossy structure are completely absorbed which leads to a decrease in lifetime. This is conventionally termed as quenching due to ``lossy surface waves" \cite{amos1997modification,{ford1984electromagnetic},{drexhage1970influence},{chance1978molecular}}. As opposed to this, the evanescent waves incident on a hyperbolic metamaterial are converted to propagating waves due to which the lifetime decrease is in fact radiative. This is analyzed in figure \ref{quenching} which shows the role of quenching for a dipole emitter placed at $d=3$ nm from various substrates (thin metal, thick metal, effective medium metamaterial and multilayer system). The smooth peak at large wavevectors (figure \ref{quenching} (a) and (b)) for the thin and thick metal is indicative of a route of decay other than propagating waves or plasmons. In the effective medium approximation, the HMM supports modes with unbounded wavevectors and hence radiative modes appear inspite of the quenching. In a practical realization, metamaterial modes with ($k_x > 1/a $) do not exist and the multilayer realization shows quenching as expected.

\subsection{\it{Deviations from EMT}}
The effective medium approximation fails when the emitter is placed at a distance much closer 
than the unit cell size. This is because the interaction between the emitter and the substrate when  $d  << a$ is dominated by waves with $k_{eff} \sim (1/d) >> (1/a)$ for which the unit cell can no longer be considered subwavelength. Thus the lifetime predicted by EMT is much lower as compared to a practical realization for very small distances. The corresponding values for the lifetime between the EMT slab and the multilayer metamaterial are expressed in the squared boxes of figure \ref{quenching}. For distances of d=30 nm, there is no quenching and the 
effective medium theory results agree well with the multilayer realization (figure \ref{quenching} (c) and (d)).

\section{Experiments}
A number of experiments have been performed to test the nature of spontaneous emission from dye molecules or quantum dots near various realizations of the hyperbolic metamaterial (HMM). This paves the way for understanding the quantum properties of complex media such as HMMs. We explain many of the properties observed in experiment and comment on possible improvements.

\subsection{\it{Multilayer and nanowire plasmonic structures}}
Both the multilayer and nanowire realization of the metamaterial have been used in different wavelength ranges in experiment. Note that the nanowire achieves type I hyperbolic behavior whereas the multilayer realization leads to type II behavior in the visible range. The nanowire structure has shown up to 5 times decrease of the emission lifetime for a Rhodamine 6G dye placed above it. Likewise, quantum dots and dye molecules on a $Au$/$Al_2O_3$ and $Ag$/$TiO_2$ multilayers have shown a decrease of lifetime by around 1.5-2. This large discrepancy in experiment can be attributed to the fact that the wavelength at which $\epsilon_{zz} \approx 0$ for the nanowire structure lies within emission spectrum of the dye molecules. This leads to an additional enhancement due to a plasmonic resonance effect.
The absorption and emission spectrum of the dye molecules are also altered in the vicinity of the metamaterial from that in solution. Since dye molecules are often suspended in a thin epoxy (dielectric) layer, light trapping causes enhanced absorption. Highly reflective type II metamaterials can lead to better absorption. This does not affect the lifetime but can affect the net output recorded at the detector \cite{sun2007practicable}.

\subsection{\it{Intrinsic quantum yield}}
The intrinsic quantum yield takes into account the intrinsic non-radiative decay of fluorophores on excitation. The intrinsic quantum yield is assumed to be independent of distance as long as there are spacer layers separating the molecules from the metal layer. The spacer layer prohibits charge transfer processes which cannot be accounted for by the theory outlined above.  The intrinsic quantum yield of dye molecules is low when compared to quantum dots thus the coupling with the metamaterial is reduced and the change in the lifetime is not significant (see figure \ref{quenching} insets). 

\subsection{\it{Apparent quantum yield and roughness}}
The apparent quantum yield is the ratio of the light collected in the far-field  by the detector compared to the case when the emitter is in vacuum.  The experiments were performed with the excitation source and detector on the same side of the metamaterial. In the ideal case, since all of the emission in the near field goes into high-\emph{k} states which cannot reach the far-field, we expect the apparent quantum yield to be significantly decreased \cite{chance1978molecular,{gontijo1999coupling}}. Note that this is different from quenching where the lifetime and intensity decrease is due to the electromagnetic energy going irreversibly into heat.

Roughness in the metamaterial structure induces major deviations in the observed apparent quantum yield. The metamaterial states can be out-coupled to propagating waves in vacuum due to roughness and non-uniformities which are inevitable in thin multilayer stacks. We note that the apparent quantum yield for a type II metamaterial which has high reflectivity can be increased through the surface roughness-assisted out-coupling of the metamaterial states. One important observation is that the apparent quantum yield will be very low for type I metamaterials because of the low reflectivity and high transmission.

\subsection{\it{Distributed dipole model}}
To compare the experimentally measured lifetime with the semiclassical theory, the system can be modeled as consisting of many dipoles (oriented randomly) interacting with the metamaterial (or metal) substrate. These dipoles embedded inside the thin dielectric matrix (thickness D) interact with the substrate and part of the reflected light is collected in the far field. The variation of emission intensity with time can be obtained as $I(t)=\sum_{0<d<D}I(d)\exp(-\Gamma(d)t)$
where $I(d)$ is the reflected intensity in the far field due to a dipole placed a distance $d$ away from the substrate and $\Gamma(d)$ is its lifetime \cite{jacob2010engineering}.

\subsection{\it{Finite sized emitters}}
One limitation of the theoretical treatment is the assumption that the emitters are significantly subwavelength and hence can be represented by point electric dipoles. While this assumption works very well for dye molecules such as $Eu^{+3}$ ions \cite{amos1997modification}, it fails for quantum dots which can be at least 3-5 nm in size.  In this case, the size of the emitter itself cannot be ignored and leads to deviations from the theoretical prediction. We note that a finite-sized emitter leads to an upper cut-off in the wavevector of the emitted waves. One expects this to curtail the coupling of high-\emph{k} waves and hence reduce the maximum possible enhancement as well as the in-coupled power \cite{poddubny2011spontaneous}. But in complete contradiction, the opposite would occur in practice \cite{gontijo1999coupling}. The reason for that is in the presence of losses, the finite size of the emitter imposes a large wavevector cut-off which actually precludes coupling to detrimental lossy surface waves. Thus a finite-sized emitter even in extremely close proximity decays radiatively, emitting a large fraction of the power into plasmonic and metamaterial modes. For careful optimization one has to take into account the wavevector cut-off due to three length scales which are the size of the emitter, size of the unit cell and the distance of the emitter from the metamaterial.

\section{Future}\label{future}
\subsection{\it{Broadband Purcell Effect}}
Future work will mainly be concentrated on out-coupling the metamaterial states using appropriate grating structures. To conclusively separate the effect of the metamaterial states from non-radiative paths, an experiment in a transmission mode will be needed. The  non-radiative and radiative decay routes can then be ascertained since the light has to pass through the metamaterial and reach the far field before being detected. Both time resolved and time integrated (spectral) measurements are necessary to identify the predicted broadband Purcell effect \cite{jacob2009broadband}. Purcell factors of $F_p=50$ over a broad bandwidth should be attainable by current fabrication methods. One of the first applications of hyperbolic metamaterial based quantum nanophotonics will be single photon sources. The broadband response, tunability and ease of fabrication make it ideally suitable for emitters such as nitrogen vacancy centers in diamond.

\subsection{\it{Active metamaterials}}
The lifetime of molecules and quantum dots are significantly decreased once embedded inside the metamaterial. This is because of the strong overlap between the metamaterial modes and the emitters. The emitters within the dielectric layers comprising the hyperbolic metamaterial can also provide gain to compensate for the losses. Active HMMs are expected to improve the device performance in all applications from imaging and nanowaveguiding to nonlinear optics and sensing \cite{tumkur2011control,ni2011loss-compensated}.

\section{Conclusion}
To summarize, we presented a comprehensive review of the design and characterization procedures for multilayer and nanowire HMMs. They are expected to be among the most promising metamaterials in the optical wavelength ranges with multitude of applications from imaging and sensing to quantum optics. Recent progress in the quantum applications of hyperbolic metamaterials and the corresponding experimental results were also reviewed.
\section{Acknowledgements}
Z. Jacob wishes to thank E. E. Narimanov for many stimulating discussions. This work was supported in part by funds from the University of Alberta start-up, National Science and Engineering Research Council of Canada, Canadian School of Energy and Environment, Nanobridge, Alberta Innovates scholarship and Queen Elizabeth II scholarship.

\bibliographystyle{unsrt}
\bibliography{QNP_HMM}
\pagebreak

\begin{appendix}
\section*{Appendix}
\section{Effective medium theory for multilayer structures}
We provide here the additional details for the theoretical model of an effective medium. As mentioned, the extreme anisotropic response of hyperbolic metamaterials can be realized for a multilayer system with alternating metal and dielectric layers, where the relative permittivities are given by $\epsilon_m$ and $\epsilon_d$ respectively.

To achieve homogeneity, the thickness of the layers must be much smaller than the operating wavelength. In practice, we require the layer thicknesses to be at least smaller than $\lambda/10$ for EMT to be valid. In this regime, the uniaxial components of the dielectric tensor, $\epsilon_{||}$ and $\epsilon_{\perp}$, are given by
\begin{equation}
\epsilon_{||} = \rho \epsilon_m + (1-\rho)\epsilon_d 
\end{equation}
and
\begin{equation}
\epsilon_{\perp} = \frac{\epsilon_m\epsilon_d}{ \rho \epsilon_d + (1-\rho)\epsilon_m},
\end{equation}
where $\rho$ is the fill-fraction of the metal in the unit cell, i.e. $\rho=\frac{t_m}{t_m+t_d}$ for the given metal and dielectric thickness $t_m$ and $t_d$. Accordingly, the plasmon resonance given by $\epsilon_{||}$ is shifted to different wavelengths depending on the fill-fraction of the metal. 

\section{Effective medium theory for cylindrical nanowires}
Following a similar approach, we can find an equivalent effective medium representation of the nanowire geometry given in figure \ref{materials} (c) by using the generalized Maxwell-Garnett approach. Assuming that we have an isotropic distribution of cylindrical nanorods, the permittivity tensor components are given by
\begin{equation}
\epsilon_{\perp} = \rho \epsilon_m + (1-\rho)\epsilon_d 
\end{equation}
and
\begin{equation}
\epsilon_{||} = \frac{(1+\rho)\epsilon_m\epsilon_d+(1-\rho)\epsilon_d^2}{(1-\rho) \epsilon_m + (1+\rho)\epsilon_d}.
\end{equation}
The axis of the cylinder nanowires is aligned along the $z$, or perpendicular, direction. These expressions are found under the assumption that $\rho<<1$ such that the local electric field across the cross-section of the nanowire is homogeneous\cite{elser2006nanowire}. The small fill-fractions used in the nanowire system generally results in much lower losses in the metamaterial system since less metal is used overall. 

\section{Dyadic Green's function formalism}
To capture the near-field effects in nanophotonic structures, we use the dyadic Green's function formalism. The Green's tensor is related to the radiation field produced by an oscillating electric dipole source in free space. The result, expanded as a summation of plane waves using the Weyl identity, leads to
\begin{equation}
\overleftrightarrow{G}=\frac{i}{8\pi^2}\int\int \frac{dk_x dk_y }{k_z k_1^2}e^{i(k_xx+k_yy+k_z|z|)} 
\left( {\begin{array}{ccc}
k_1^2-k_x^2&-k_xk_y&\mp k_xk_z\\
-k_yk_x&k_1^2-k_y^2&\mp k_yk_z\\
\mp k_zk_x&\mp k_zk_y&k_1^2-k_z^2
 \end{array} } \right)
\end{equation}
where $k_1=\sqrt{\epsilon_1}\omega/c$ is the angular wavevector of a medium with permittivity $\epsilon_1$. We can decompose the Green's function tensor into its (s)- and (p)-polarized components $\overleftrightarrow{G}_s$ and $\overleftrightarrow{G}_p$, such that
\begin{eqnarray}
 \overleftrightarrow{G}_s=\frac{i}{8\pi^2}\int\int \frac{dk_x dk_y }{k_z(k_x^2+k_y^2)}e^{i(k_xx+k_yy+k_z|z|)}  A_s \\
\overleftrightarrow{G}_p=\frac{i}{8\pi^2}\int\int\frac{dk_x dk_y }{k_z (k_x^2+k_y^2) k_1^2}e^{i(k_xx+k_yy+k_z|z|)} A_p
\end{eqnarray}
where $A_s$ and $A_p$ are the tensors
\begin{eqnarray}
A_s = 
\left( {\begin{array}{ccc}
k_y^2&-k_xk_y& 0\\
-k_yk_x&k_x^2&0\\
0&0&0
\end{array} } \right), \\
A_p = 
\left( {\begin{array}{ccc}
k_x^2k_z^2&k_xk_yk_z^2& \mp k_xk_z(k_x^2+k_y^2)\\
k_xk_yk_z^2&k_y^2k_z^2& \mp k_yk_z(k_x^2+k_y^2)\\
 \mp k_xk_z(k_x^2+k_y^2)& \mp k_yk_z(k_x^2+k_y^2)&(k_x^2+k_y^2)^2
 \end{array} } \right).
\label{GreenTensor}
\end{eqnarray}
Next, we formalize the expression of the decay rate $\Gamma$ by following the theory developed by Ford and Weber \cite{ford1984electromagnetic}. This approach yields the normalized decay rate, in terms of the wavevector-resolved local density of states, such that
\begin{equation}
\frac{\Gamma}{\Gamma_o} = 1 -\eta+ \eta\, \mathrm{Re}\int\limits_{0}^{~\infty} \rho(\omega,d,\vec{k}) dk_x
\label{lifetime}
\end{equation}
where $\eta$ is the intrinsic quantum yield of the emitter and $\Gamma_o$ is the decay rate of the emitter in free-space. The wavevector-resolved local density of states, normalized to free-space, is then defined as
\begin{equation}
\rho(\omega,d,\vec{k})=  \frac{3}{2 |p|^2k_1^3} \frac{ k_x }{k_z}\left\{ \frac{1}{2} p_{||}^2 [(1+r^s\mathrm{e}^{2ik_zd})k_1^2+(1-r^p\mathrm{e}^{2ik_zd})k_z^2]+p_\perp^2 (1+r^p\mathrm{e}^{2ik_zd})k_x^2 \right\}
\label{wavevector_ldos}
\end{equation}
where $p_{||}$ and $p_\perp$ are the parallel and perpendicular components of the dipole moment, $d$ is the distance of the dipole from the planar interface, and $r^{s,p}$ is the reflection coefficient for (s)- and (p)- polarized light respectively. 

The reflection coefficients can be found by using the Fresnel equations for simple configurations, however, the use of the transfer matrix method is required to obtain the necessary values in multilayered structures. Considering a system of $N$ total layers, the general expression is
\begin{equation}
\left( {\begin{array}{c}
1 \\r^{s,p} 
\end{array} } \right) 
=(D_0^{s,p})^{-1} T^{s,p} D_{N+1}^{s,p}
\left( {\begin{array}{c}
t^{s,p} \\0 
\end{array} } \right) 
\end{equation}
where
\begin{eqnarray}
T^{s,p}=\left(\prod_{i=1}^N D_{i}^{s,p}P_i (D_{i}^{s,p})^{-1}\right), \\
 D_i^p=
\left( {\begin{array}{cc}
1&1\\
\frac{k_z^i}{\epsilon_i}&-\frac{k_z^i}{\epsilon_i}\\
\end{array} } \right) 
, D_i^s=
\left( {\begin{array}{cc}
1&1\\
k_z^i&-k_z^i\\
\end{array} } \right) 
, P_i=
\left( {\begin{array}{cc}
 e^{-j k_z^i d_i}&0\\
0& e^{j k_z^i d_i}\\
\end{array} } \right).
\end{eqnarray}
The multilayer system consists of layers with dielectric constant $\epsilon_i$ and thickness $d_i$, while $k_z^i$ is the perpendicular wavevector component inside the $i$th layer. Combining all of these results, we can calculate W-LDOS expression in equation (\ref{wavevector_ldos}), while using numerical integration techniques to evaluate the normalized decay rates in equation (\ref{lifetime}).

\end{appendix}

\end{document}